\documentclass[graybox]{svmult}

\usepackage{mathptmx}       
\usepackage{helvet}         
\usepackage{courier}        
\usepackage{type1cm}        
\usepackage{makeidx}         
\usepackage{graphicx}        
\usepackage{multicol}        
\usepackage[bottom]{footmisc}
\usepackage{amsmath}
\usepackage{caption}
\usepackage{wrapfig}
\usepackage{url}
\usepackage{epsfig}
\usepackage{subcaption}
\usepackage{multirow}
\usepackage{subfig}
\usepackage{tabularx}

 \newcommand{\sgn}{\mathop{\mathrm{sgn}}}

\makeindex             

\begin{document}
\title*{Stochastic modeling of cargo transport by teams of molecular motors}
\author{Sarah Klein, C\'ecile Appert-Rolland and Ludger Santen}
\institute{Sarah Klein \at Universit\"at des Saarlandes , Postfach 151150 - Geb\"aude E2.6 , 66041 Saarbr\"ucken, Germany ; LPT, Batiment 210, University Paris-Sud,
F-91405 ORSAY Cedex - France \ \email{sarah.klein@th.u-psud.fr}
\and C\'ecile Appert-Rolland \at LPT, Batiment 210, University Paris-Sud,
F-91405 ORSAY Cedex - France 
\and Ludger Santen \at Universit\"at des Saarlandes , Postfach 151150 - Geb\"aude E2.6 , 66041 Saarbr\"ucken, Germany}
%
%
%
\maketitle

\abstract{Many different types of cellular cargos are transported bidirectionally along microtubules by teams of molecular motors. 
The motion of this cargo-motors system has been experimentally characterized \textit{in vivo} as processive with rather persistent directionality.
Different theoretical approaches have been suggested in order to
explore the origin of this kind of motion. An effective theoretical approach, 
introduced by M\"uller \textit{et al}. \cite{Mueller2008}, describes the cargo dynamics
as a tug-of-war between different kinds of motors.  An alternative approach has been 
suggested recently by Kunwar  \textit{et al}. \cite{Kunwar}, 
who considered the coupling between motor and cargo in more detail. \\
Based on this framework we introduce a model considering single motor positions which we propagate in continuous time. Furthermore, we analyze 
the possible influence of the discrete time update schemes used in previous publications 
on the system's dynamic. }

\section{Introduction}
\label{intro}

In the last years bidirectional motion along microtubules was observed in many different cell types \cite{Welte2004,Steinberg2004}. 
In most of these cells it is still not clear how this bidirectional motion is realized. 

Similar to a human road network connecting different places, the cell provides several
filaments which can be used for directed transport. Besides the transport utility, the
filaments give the cell its characteristic shape. To achieve this double goal the cell produces a cortex
of filaments close to the membrane and radial growing filaments from the nucleus to
the periphery. The 
set of these filaments constitutes the cytoskeleton. Intracellular
transport along microtubules, which is a radial growing filament, is managed by
mainly two kinds of transporters, the so-called molecular motors which are identified
as kinesin and dynein \cite{Hirokawa}. The principal difference between the two kinds of motors
is their preferred walking direction. The microtubules are polarized, i.e. they have
well-defined directions which are called plus- and minus-direction, respectively.
Kinesin’s preferred orientation is to the plus-end of the microtubule, while dynein’s
orientation is opposed. Assuming that microtubules mainly grow with their plus end
to the cell periphery cellular cargos can be moved to the nucleus and to the membrane
by dynein and kinesin, respectively. However, secretory cargos which could be thought to
leave the cell as fast and straight as possible, actually show a saltatory motion \textit{in vivo} \cite{Trinczek}. This
behavior suggests that a number of cellular cargos exists, on which
kinesins as well as dyneins are bound at the same time. One possible reason for this
motion is to pass obstacles by a second try \cite{Dohner}. The detailed mechanisms,
which are leading to this unconventional bidirectional motion are for most of
the motor-cargo systems still unknown. 

To describe this bidirectional motion theoretically two mechanisms have been suggested: The first
one assumes that $N_+$ kinesins and $N_-$ dyneins are involved in a mechanical tug-of-war and
fight for the direction the cargo effectively moves, while the second one requires a
control mechanism to achieve coordinated \textit{in vivo}-behavior \cite{Gross2002,Welte1998}.
The \textit{pure} tug-of-war model was introduced by M\"uller  \textit{et al}. \cite{Mueller2008} to describe lipid droplet movement in
evolving \textit{Drosophila} embryo cells. They use a mean-field model, meaning that the motors of one team share the load equally.
As a consequence, all kinetic quantities are determined
by the number of attached motors to the filament.
Indeed, since the motors can bind to and unbind from the filament, the number of motors of each kind attached to the filament
fluctuates between zero and $N_\pm$ with time.
Between two attachment/detachment events, the cargo's
velocity is constant and 
determined by the strength
of the two teams (which depends on the number of attached
motors).
The number of attached motors also determines
the load force felt by each team of motors,
and exerted by the opposite team via the cargo.
Once one motor detaches one observes a cascade of
detachments of motors of this kind and therewith
it is possible, in the framework of this model, 
to generate motility states with high velocity,
where one team {\em wins} over the other.

This model is quite elegant since experimental observables
like the cargo's velocity can be calculated analytically.
However, since this walking in concert was not yet observed \textit{in
vitro}, Kunwar \textit{et al.} had a closer look at different
observables, like the pausing time and run length of single
trajectories but did not find matching results in
experiments. Therefore they introduced a model taking
explicitly the motor positions into account and models the
motor-cargo coupling as a linear spring.

In this contribution, we introduce a general model
with simple reaction rates
which propagates the cargo along its
equation of motion in {\em continuous} time. Furthermore, we
compare and discuss the consequences of using different
update schemes.
\section{Model }\label{model}
\begin{figure}[bt]
\begin{minipage}[c]{0.65\textwidth}
\centering
\includegraphics[scale=0.25]{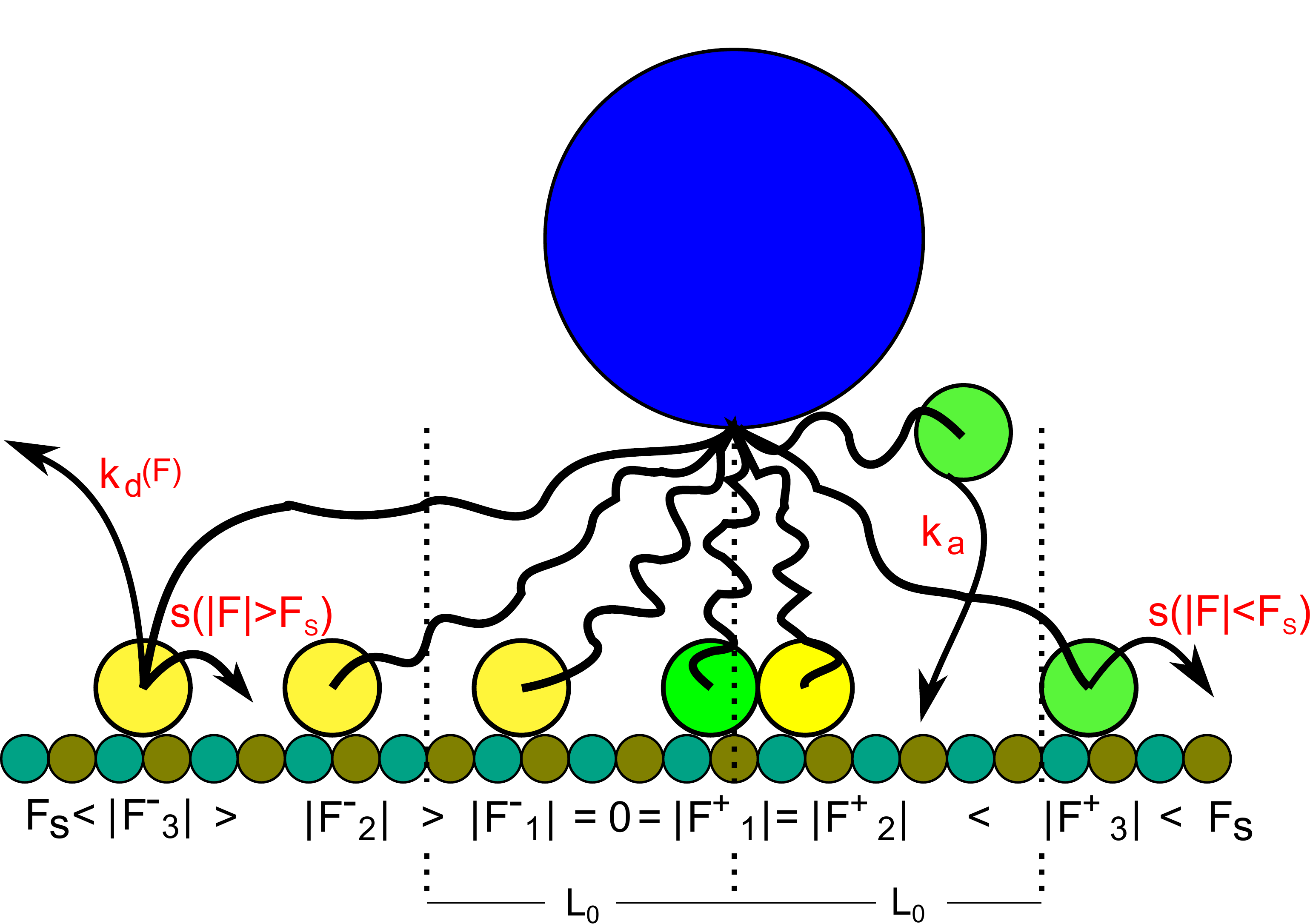}
\caption{Sketch of the model dynamics. Two kind of motors:
one team prefers to walk to the plus-end (green) while the
other prefers walking to the minus-end (yellow) of a microtubule.
Within the region $L_0$ around the cargo's center of
mass the motors apply no force on the cargo (blue).} 
\label{skizze}
\end{minipage}\ \ \ \ \ \ \begin{minipage}[c]{0.25\textwidth}
\renewcommand{\figurename}{Table}
 \addtocounter{figure}{-1}
\flushright 
\begin{tabular}{|c|c|} 
\hline
$N_\pm$ & 5 \\ \hline
 $v_{f}$ & {$1000 \ \textnormal{nm/s} $}  \\ \hline
$v_{b}$ & 6 nm/s \\ \hline
$D$ & 0.32 pN/nm \\ \hline
$F_D$ & 3 pN \\ \hline
 $k_a$ & 5.0  s$^{-1}$ \\ \hline
 $k_d^0$ & {$1\ {\text{s}}^{-1}$}  \\ \hline
$F_S$ & 6 / 2 pN  \\ \hline 
$R$ & 250 nm \\ \hline
\end{tabular}
\caption{Simulation parameter for the results below.}
\label{paraset}
\end{minipage}
\end{figure}
\renewcommand{\figurename}{Fig.}
Inspired by the bidirectional cargo transport models of \cite{Mueller2008,Kunwar} we define a stochastic model to move a cargo by teams of molecular motors along a microtubule.  $N_+$ and $N_-$ motors are tightly bound to the cargo and pull it in plus- and minus-direction, respectively. 
In contrast to \cite{Mueller2008} and in agreement with \cite{Kunwar} we take every single motor position $x_i$ into account and calculate the thereby generated force $F_i$ on the cargo. 
We model the motor tail, which permanently connects the motor head to the cargo, as linear spring with an untensioned length $L_0$ and a spring constant $\alpha$. 
In contrast to M\"uller's model \cite{Mueller2008} where the motors can attach (with rate $k_a$) to and detach from the filament (with force-dependent rate $k_d(F_i)$) only, 
in our model the motors can once bound to the filament, a one-dimensional infinite lattice, can make a step of size $d$ with a force-dependent rate $s(F_i)$. 
Since it seems to be biological relevant that the motors feel no force when they attach to the filament we 
reduce the allowed attachment region to $\pm L_0$ around the center of mass of the cargo.

Due to the de-/attaching events the number $n_\pm$ of plus (minus) motors bound to the filament is in the range $0 \leq n_+ \leq N_+$ ($0 \leq n_- \leq N_-$). 
The resulting force on the cargo at position $x_C(t)$ at time $t$ is then given by the sum of all single forces
\begin{eqnarray}
&F& (x_C(t),\{x_i\})= \sum_{i=1}^{n_++n_-}  F_i (x_C(t),\{x_i\}) \\
&=& \sum_{i=1}^{n_++n_-}  \alpha\bigg[ \Big( (x_i-x_C(t))-L_0 \sgn (x_i-x_C(t)) \Big) \cdot \nonumber \Theta \big(|x_C(t)-x_i|-L_0\big) \bigg] \nonumber,
\end{eqnarray}
with the Heaviside step function $\Theta(.)$.
In this paper we illustrate how we extend the model of Kunwar \textit{et al.} for a continuous time propagation of the cargo in the case of the simple relations for the stepping and detachment rates introduced in \cite{Mueller2008} and given below. 

The motors cannot stand arbitrarily high forces. Thus the so-called stall force $F_S$ gives the maximal force under which a motor 
can walk in its preferred direction. We split the stepping rate $s(F_i)$ in two regimes: 
(I) forces smaller in absolute value than the stall force ($|F_i|<F_S$) where the motors walk in their preferred direction and (II) forces bigger in absolute value than the stall force ($|F_i|\geq F_S$) where the motors walk opposed to their preferred direction and use
\begin{equation}
s(|F_i|)= \begin{cases} 
\frac{v_f}{d} \Big(1-\frac{|F_i|}{F_S} \Big) \ \ \ \ |F_i|<F_S \\	\\	
\frac{v_b}{d}\Big(1-\frac{|F_i|}{F_S}\Big) \ \ \ \ |F_i|<F_S \end{cases} 
\end{equation}
with $v_b \ll v_f $ \cite{Carter,Mallik2005}.\\
Assuming that the motors can walk on several close microtubules in a crowded environment and that their attachment point to the cargo is not necessarily the same, a sterical exclusion of the motor heads on the lattice is not regarded in the model. 

For the detachment rate we use \cite{Mueller2008}
\begin{equation}
k_d(|F_i|) = k_d^0 \exp\left(\frac{|F_i|}{F_D}\right), 
\end{equation} 
with the force-free detachment rate $k_d^0$ and the detachment force $F_D$, which determines the force scale.\\

{\bf Update mechanisms}\\

In the mean-field model \cite{Mueller2008} the cargo moves with a constant velocity during two motor events, calculated by the number of attached motors of each team. 
The time at which the next event occurs, is calculated by means of Gillespie's algorithm \cite{Gillespie}. Within this framework the cargo's velocity is piecewise linear. \\
Kunwar \textit{et al}. \cite{Kunwar} use a parallel, thus discrete time update scheme to propagate the system. 
At every fixed time step $\Delta t$ they calculate the probability that a motor event occurs within this time step. An event should be rare within $\Delta t$ to get a good approximation of the the exact solution in continuous time. In their simulations they use $ \Delta t = 10^{-6}$ s. \\ 
Once the motor dynamic is determined, one has to decide how the cargo reacts to each change in the 
motor configuration. In \cite{Kunwar} two different cargo
dynamics are introduced:
either the cargo moves instantaneously 
to the position with balanced forces,
or it 
undergoes a viscous force from the environment.
The mean-field model of \cite{Mueller2008} was treated in
the case of 
an instantaneously reacting cargo.

In \cite{Kunwar} 
a viscous environment was taken into account
by calculating the 
position of a cargo with radius $R$ after $\Delta t$ according to
\begin{equation}
 x_C(t+\Delta t) = x_C(t) + \sum_{i=1}^{n_++n_-} \frac{F_i}{6\pi\eta R},
\end{equation}
where $\eta$ is the fluid's viscosity.

To get a more general approach we rather use the cargo's equation of motion
\begin{equation}
m \frac{\partial^2 x_C(t) }{\partial t^2 } = -\beta \frac{\partial x_C(t)}{\partial t} + \sum_{i=1}^{n_++n_-}F_i(x_C(t),\{x_i\}) ,
\label{eqofm_supp}
\end{equation}
with $\beta = 6\pi\eta R$ and the cargo's mass $m$, to determine the time-dependent 
position of the cargo.

By determining the force applied on each motor by the distance between motor head position and the center of mass of the cargo, the force $F_i$  depends on time, too. 
Hence, the motor rates for stepping and detaching are time-dependent. Thus the cargo moves in a viscous medium in a harmonic potential of the sum of the springs. 
Note that the number of engaged springs changes, if the distance between a motor and the cargo falls below or exceeds $L_0$. 
Therefore we have to solve eq. (\ref{eqofm_supp}) piecewise on segments with a constant number of motors which pull the cargo.
On every single segment we solve the equation
\begin{equation}
m \frac{\partial^2 x_C(t) }{\partial t^2 } = -\beta \frac{\partial x_C(t)}{\partial t} -\epsilon x_C(t) + \epsilon\zeta,
\label{eqofm_red}
\end{equation}
with 
\begin{equation}
\epsilon = \sum_{i=1}^{n_++n_-} \alpha\cdot \Theta \big(|x_C(t)-x_i|-L_0 \big) 
\end{equation}
which determines the effective spring constant and
\begin{equation}
\zeta =\sum_{i=1}^{n_++n_-}	\big(x_i-\sgn(x_C(t)-x_i)L_0\big) \cdot\Theta \big(|x_C(t)-x_i|-L_0 \big) ,
\end{equation}
the effective potential generated by the given motor configuration. We then get the cargo position $x_C(t)$ at time $t$ on the segments with constant number of pulling motors
\begin{equation}\label{x_von_t}
x_C(t)=\frac{\lambda_1\zeta+\lambda_1x_0-v_0}{\lambda_1-\lambda_2} \exp(\lambda_2 t) + \frac{v_0-\lambda_2\zeta-\lambda_2x_0}{\lambda_1-\lambda_2} \exp(\lambda_1t) - \zeta
\end{equation}
with
\begin{equation}
\lambda_1= -\frac{\beta}{2m}+\sqrt{\left(\frac{\beta}{2m}\right)^2-\frac{\epsilon}{m}} \ \ \ \ \text{ and }  \ \ 
\lambda_2=-\frac{\beta}{2m}-\sqrt{\left(\frac{\beta}{2m}\right)^2-\frac{\epsilon}{m}}.
\end{equation}
Now knowing the cargo position at an arbitrary time $t$ we can use Gillespie's algorithm for time-dependent rates \cite{Gillespie_1978} to calculate the next event time.
%
%
\section{Results}
\begin{figure}[b]
\centering \includegraphics[scale=0.23]{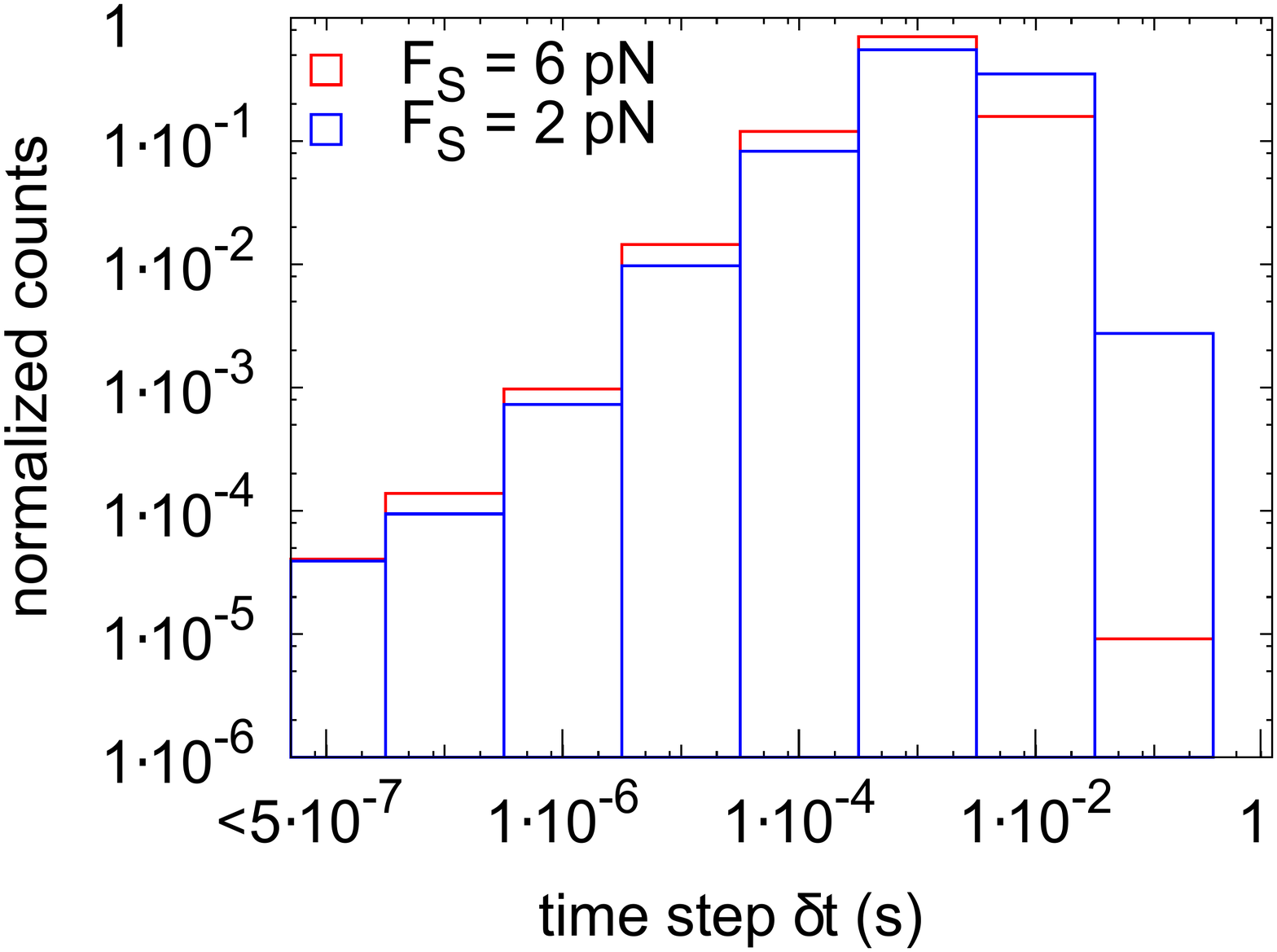}
\sidecaption
\caption{Log-log plot of the normalized count of times between two occurring events calculated by the exact algorithm \cite{Gillespie_1978} for $F_S = 2$ pN (blue) 
and $F_S=6$ pN (red). Obviously, times between events smaller than $10^{-6}$ s occur for both stall forces. }
\label{hvelo}       
\end{figure}
At first we analyze the distribution of times between two motor updates which we generate with the exact algorithm and the parameter set given in Table \ref{paraset}.

In Fig. \ref{hvelo} the normalized count of times between events is shown in a double logarithmic plot.
Obviously, times smaller than $\Delta t=10^{-6}$ s occur if we propagate the system with the exact algorithm. By analyzing $10^{5}$ events we calculated the mean time between events $\langle t \rangle$ for the two stall forces as well as the smallest $t_s$ and the longest time $t_l$ between two events and get
\begin{eqnarray*}
&\text{\underline{$F_S = 2$ pN}}& \ \ \ \langle t \rangle =3.5\cdot 10^{-3} \text{ s}  \ \ \ t_s = 3.1\cdot 10^{-8} \text{ s} \ \ \  t_l =  1.8\cdot 10^{-1} \text{ s} \\
&\text{\underline{$F_S = 6$ pN}}& \ \ \ \langle t \rangle =1.8\cdot 10^{-3} \text{ s}  \ \ \ t_s = 1.6\cdot 10^{-8} \text{ s} \ \ \  t_l =  8.1\cdot 10^{-2} \text{ s}. 
\end{eqnarray*}
It remains the question how this influences the system's observables.
In \cite{Kunwar} the focus is on the run length and pause
duration of the single walks.
However, as the motion is stepwise, these observables are defined
from quite arbitrary time/distance thresholds.
That is why we prefered to concentrate on another
quantity to
compare our data to the parallel update
scheme, namely the discrete velocity
\begin{equation}
\tilde{v} = \frac{|x(t+\text{D}t) - x(t)|}{\text{D}t},
\end{equation}
where we use $\text{D} t = 0.16$ s as it was suggested in \cite{Mueller2008}. \\
\begin{figure}[tb]
\begin{minipage}{0.5\textwidth}
\hspace*{-0.2cm} \includegraphics[scale=0.23]{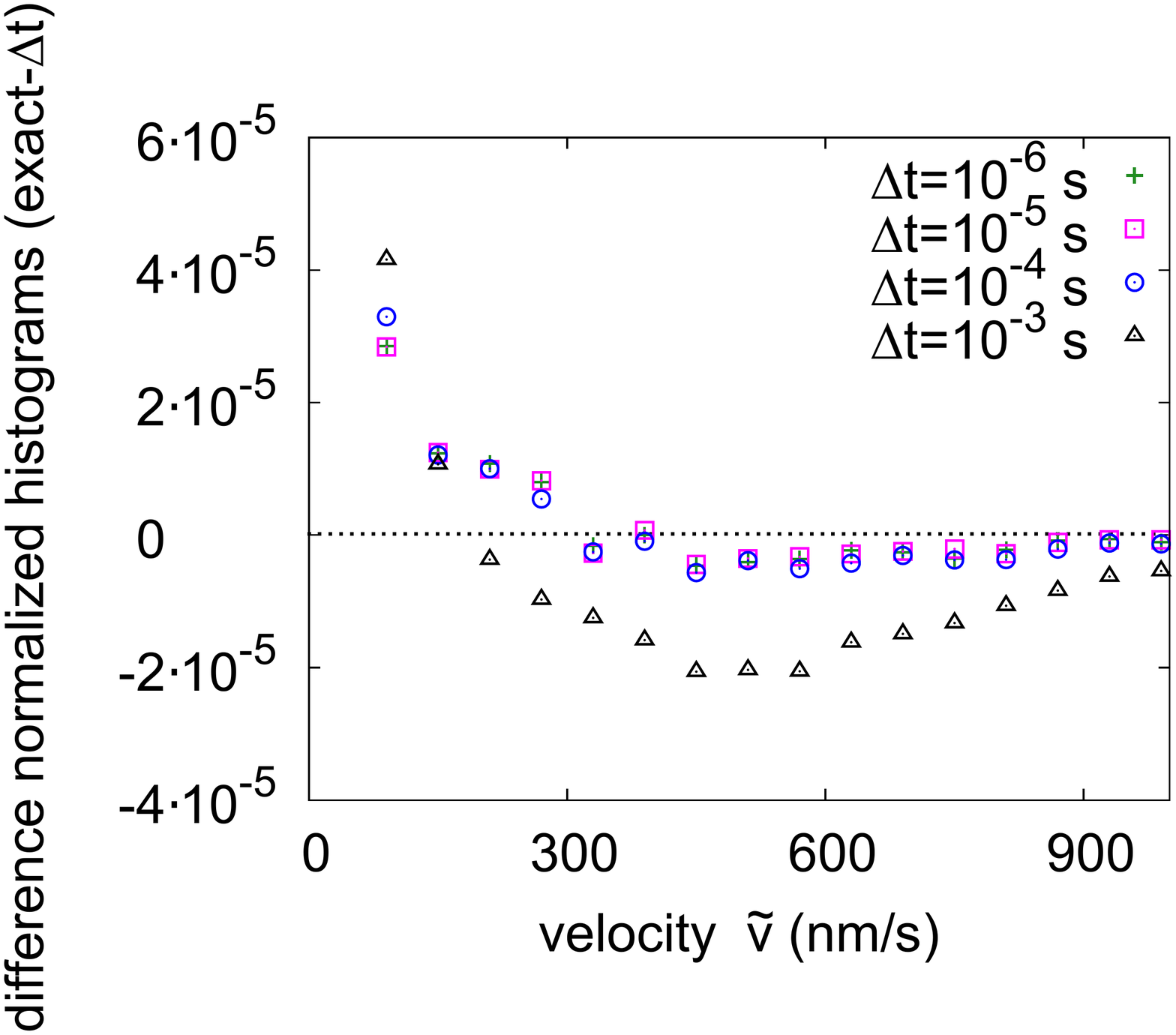}
\subcaption{}
\end{minipage}
\begin{minipage}{0.5\textwidth}
\hspace*{-0.2cm} \includegraphics[scale=0.23]{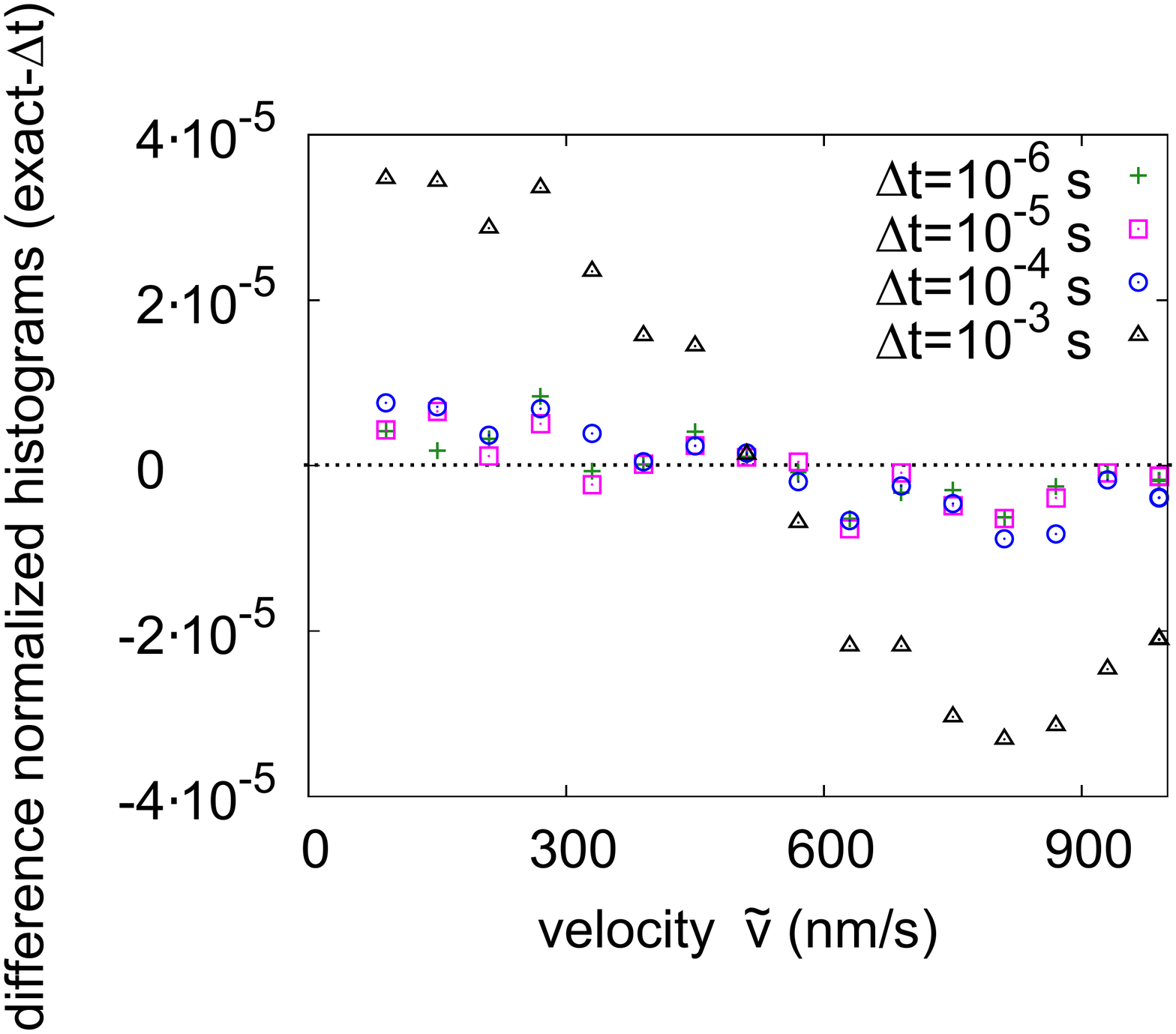}
\subcaption{}
\end{minipage}
\caption{Difference between the exact velocity histogram generated with Gillespie's algorithm \cite{Gillespie_1978} and the parallel update scheme suggested in \cite{Kunwar} with different $\Delta t$ for \textbf{(a)} $F_S=2$ pN and \textbf{(b)} $F_S=6$ pN. Sample size per histogram: $5\cdot 10^{6}$.}
\label{hdiff}       
\end{figure}
In Fig. \ref{hdiff} we show the difference between the normalized velocity histogram generated by Gillespie's algorithm \cite{Gillespie_1978} and those generated by the parallel update scheme for different $\Delta t$ and for two different stall forces $F_S$. 
In both cases an increase in $\Delta t$ increases the cargo's velocity as shown in Table \ref{tab_vel}. \\
%
 \addtocounter{table}{1}
\begin{table}[tbh]
\centering\begin{tabularx}{0.9\textwidth}{ |X|X|X|X|X|X| }
\hline
 &  &\multicolumn{4}{c|}{$\Delta t$} \\
$F_S$ & exact & $10^{-6}$ s & $10^{-5}$ s& $10^{-4}$ s& $10^{-3}$ s \\ \hline \hline
\multirow{2}{*}{2 pN} &149.0 &149.8 & 149.7 &150.3 & 155.4\\ 
	&				 $\pm$0.10 &	$\pm$ 0.05 &$\pm$  0.05 &$\pm$ 0.08 & $\pm$ 0.08 \\ \hline 
\multirow{2}{*}{6 pN} & 449.4  & 450.2 &  450.4 &451.1 & 458.5 \\ 	 	
&$\pm$0.18 &$\pm$0.09 &$\pm$0.13 &$\pm$0.13 &$\pm$0.13 \\ \hline
\end{tabularx}
\caption{Mean discrete velocity $\langle \tilde{v} \rangle $ in nm/s for the exact and the parallel update. } 
\label{tab_vel}
\end{table}
\section{Discussion}
We have introduced in this contribution an exact algorithm to propagate the motors-cargo system in continuous time. 

An analysis of the times between two events
reveals that very different time scales are involved:
while most  times between two events are greater than $10^{-3}$ s,
a fraction of events are separated by less than $10^{-7}$ s.

From our results, a first conclusion is that if one uses parallel update,
the time step $\Delta t$ should at least be less than
$10^{-5}$ s
to expect results in good agreement with the continuous
time dynamics.
However, the continuous time dynamics reveals that much
shorter time scales are involved, as a signature of cascades
of events. These cascades are overlooked in the discrete
updates even
for time steps as small as $\Delta t=10^{-6}$ s.
While we have found that this approximation does not alter the quality
of measurements of most quantities when such a small
time step is used (as it is the case in 
\cite{Kunwar}), one cannot exclude that for some
other sets of parameters, and/or for more sensitive
observables, these cascades could have a stronger
effect.
Actually, though further numerical support should be
provided to conclude, our results seem to indicate that
discrete updates systematically slightly underestimate the
probability to have weak cargo velocities (unless
prohibitively small time steps would be used).
This can be understood as an effect of the synchronization
of the motors induced by the time discretization, 
similarly to what happens with the mean-field assumption
used in \cite{Mueller2008}
(which can also
be seen as a synchronization mechanism)
which overemphasizes large velocity states.
As a conclusion, in such a system involving very different
time scales, an exact algorithm in continuous time provides
an efficient numerical scheme: it allows to avoid any
possible artefact that would come from the discretization,
without any extra numerical cost.

In further work we will extend this model to more realistic motor rates and show for biologically
relevant parameter sets how some external quantity like the ATP concentration or the viscosity of the surrounded fluid can control the drift of the cargo~\cite{Klein2014}.
\begin{acknowledgement}
This work was supported by the Deutsche 
Forschungsgemeinschaft (DFG) within the collaborative 
research center SFB 1027 and the research training group GRK 1276.
\end{acknowledgement}
\bibliographystyle{alpha}

\end{document}